**Title: Successful and sustainable undergraduate research in data science**

**Authors:** Audrey E. Hendricks[1,2]
[1] University of Colorado Anschutz Medical Campus, Department of Biomedical Informatics
[2] University of Colorado Denver, Department of Mathematical & Statistical Sciences

**Abstract:** Undergraduate research experiences hold many potential benefits. Students can learn about new areas opening up previously unknown paths in academia and industry. The hands-on experience often provides a deeper understanding of what science, research, and data analysis is and, importantly, is not. While numerous studies have provided information about the benefits and challenges of undergraduate research, many still find it difficult to start an undergraduate research group. Here, we provide a roadmap and resources to help faculty of all levels create and sustain an undergraduate research group in quantitative areas such as statistics, informatics, and data science. While we focus on undergraduate research in data science, many of the recommendations may be generally useful to research mentoring of all levels and other domains.

## Introduction

Undergraduate research experiences can help students engage in a scientific area to a deeper lever, learn professional skills, and try out research to assess whether being a scientist or data scientist holds interest as a career[1]. Students can also learn important truths about the nature of science, data science, and analysis including the enormity of data wrangling, the need to iterate, and the regular nature of setbacks[1,2]. Participating in research experiences has been shown to increase resilience, confidence, and persistence in science, technology, engineering, and mathematics (STEM) degrees as well as motivation to apply to graduate programs[3-5]. In addition to the benefits to undergraduates, there are benefits to the broader research team. PhD or more senior MS or undergraduate researchers can gain mentoring experience as co-mentors. Faculty get to mentor the next generation of data scientists. And, importantly, undergraduate researchers and faculty alike get to work together to advance impactful research.

Many have studied the benefits, and barriers to undergraduate research[1,2,4-8]. Additionally, there are resources and guidance for undergraduates aiming to complete research[8]. For instance, the council on Undergraduate Research (CUR, https://www.cur.org/) is especially useful providing community and information for mentees and mentors alike[6].

Despite the many resources and publications available, many can still find it difficult to get started. A large barrier to starting can be the limit of time and resources. Indeed, many of us would like to mentor for far more time than we have available. The constraint of time is highlighted when we further breakdown our total mentoring time into individual units for each trainee (**Figure 1**). Given the restriction of time, how do we ensure our mentoring time is successful and sustainable for the PI? How do we ensure the trainee time is successful and sustainable for each student?

Below, we discuss tips for recruiting, training, and retaining undergraduates in research. Our recommendations provide options for how to adequately support trainees while also making mentoring undergraduate students sustainable alongside other career expectations such as teaching, service, and, of course, research.

Although we focus on undergraduates and data science, many of our recommendations are useful for all levels of trainees and other research domains.

## Resources

Resources, such as an expectations document, meeting guidelines, progress report, and self-evaluation are provided in the **Appendix**. All documents are available for free use, modification, and reuse.

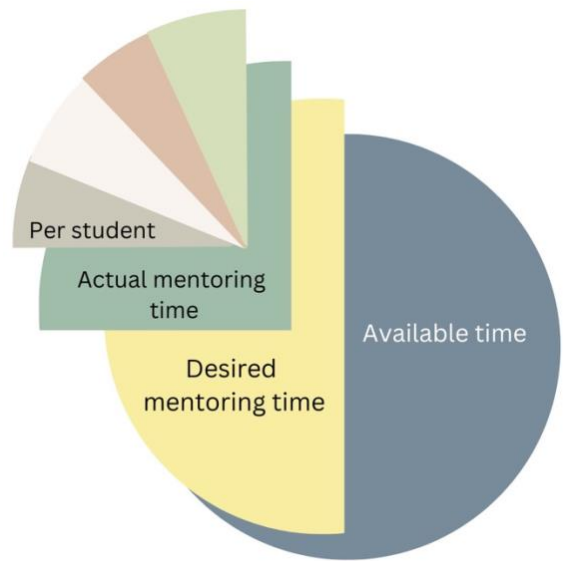

**Figure 1**. **Mentoring time pie.** Time pie showing the desired time for mentoring (yellow) is often much greater than the available time for mentoring (green). Mentoring time is further decreased when splitting between students.

## Recruiting

***Start small (but not too small).*** While starting with one undergraduate student might seem like the easiest and least time-consuming way to begin, one student may feel isolated or make slow progress. Instead, starting with two or three undergraduate researchers provides students with a pod of peers to ask questions and work with. Starting two to three undergraduates together is especially beneficial for research groups without more senior trainees to provide additional support as near-peer mentors.

***Recruit a diverse team.*** Diversity can be considered across numerous dimensions including academic major, culture, ethnicity, experience, gender, socioeconomic status (SES), and race. Building a diverse team and community can help bring many perspectives increasing the generalizability of the research and helping ensure that a broad set of research questions are being investigated. Recruiting a diverse team is supported by creating broad awareness of the research opportunity, removing barriers to application, and removing barriers to participation.

***Create broad awareness of a research opportunity.*** Creating awareness of a research opportunity can occur through classes, word of mouth from both colleagues and students, a research group website, and university or college job postings and programs. For instance, University of Colorado Denver's undergraduate research program, Education through Undergraduate Research and Creative Activities (EURēCA!), helps faculty post positions for their teams, provides funding for students, and training for mentees and mentors. Reaching out to other departments, classes, and student clubs can help to broaden your reach to students who may not otherwise know about the opportunity.

***Remove barriers (as many as possible).*** Remove barriers to entry and barriers to participation. Traditional barriers to entry include GPA, class, and major requirements. Indeed, GPA is not always a measure of a student's ability. Students who work part or full time or have other responsibilities, such as family, may have lower GPA due to less time available to contribute to school. The ramifications of this can be most severe for low-income students who often need to work and work longer hours while attending school[9]. Instead, consider requirements that are more directly inline with training success such as a commitment of time (e.g., ≥8 months and ≥12 hours/week) and enthusiasm to learn and work with others. Many other skills and information can be learned along the way.

Perhaps the largest barrier to participation in research is income. When research positions are not paid, only people who can afford to volunteer a significant proportion of time will be able to participate. Work-study, where the federal government pays for up to 75% of the salary, is a cost-effective method to pay for undergraduate research assistants who qualify. Additionally or alternatively, students can sometimes take an independent study or internship course to receive course credit while completing research training. The NSF and NIH each have numerous programs to support undergraduate research including supplements to add onto existing programs (e.g., diversity supplements for NIH R mechanisms) and research training grants to support larger programs (e.g., R25s).

## Training

***Set clear expectations for both mentor and mentee.***
Setting reasonable, clear, and accessible expectations for both mentor and mentee is helpful in having a positive research experience. Providing a written document that outlines expectations for the mentee, mentor, and research team is a great resource that can be accessed throughout the research experience as needed. Expectations can range from general working guidelines (e.g., number of hours to work per week, days in person or online, response time to emails, attendance at research team meetings) to expected group behavior (e.g., supportive, inclusive) to individual expectations. Trainees can write an expectations document at the beginning of their research experience, modify an existing document, or use the mentor's document. Regardless, the document should be written and signed at the beginning of the research experience. An example expectations document is in the **Appendix**.

***Set frequent (but not too frequent) meetings.***
Research group meetings provide an opportunity to learn from other people's research and feel like part of a larger team while 1-to-1 meetings provide more opportunities for student learning. While trainees benefit from regular meetings, too many meetings can infringe on time spent training or doing research. Using no more than 10-25% of research time for meetings is a reasonable rule of thumb to get started. Shorter, more frequent meetings, such as 20min meetings weekly rather than 45 min meetings bi-weekly, can provide students with more frequent opportunities to ask questions and check in helping to prevent stalling of research. Undergraduate researchers can also tradeoff between meeting with the faculty member and near-peer or peer mentors to provide more regular check ins and have more support within the team. Faculty drop-in research hours a few times a week can also provide another time when students can stop by to ask questions in between meetings.

**Instill organization and record keeping.**
Having a record of research activities, questions, and decisions is essential for remembering and communicating research, as well as instilling a sense of accomplishment. Physical lab notebooks are common for record keeping in wet labs and can be used for data science projects. Alternatively, electronic notebooks, such as through google docs, are easily searchable, sharable, and linkable to results and other supporting documents making them ideal record keepers for data science projects. Undergraduate research assistants can be supported in leading their 1to1 and small group meetings by having a 15-5 agenda and summary structure for their electronic notebook. A 15-5 agenda and summary takes no more than 15 minutes to write and no more than 5 minutes to read and includes sections such as the agenda, accomplishments, questions, priorities for next week. Notes can be taken in the same document during the meeting. An example is provided in the **Appendix**.

***Provide frequent opportunities for goal setting and feedback.***
Completing regular goal setting and providing feedback is critical to ensuring a successful research experience. Natural opportunities for goal setting are at the start of each semester or term and while progress reports are natural to given at the end of each semester/term. While more informal feedback should be given at weekly or bi-weekly meetings, longer meetings at the beginning and end of the semester or term can be used for a discussion of formal goal setting and feedback. In addition to progress reports from the faculty and near-peer mentors, self-evaluations completed by the trainee can help identify perceived areas of strength and weaknesses and help ensure the student is on track for meeting short and long term goals . Example progress report categories and self-evaluation questions are provided below as well as example forms in the **Appendix**.

Example progress report areas of assessment (expectations):
- General. (Ontime updates and progress reports. Clear reporting of current work and goals. Accurate tracking of time and progress. Responsive in a timely manner to emails and other communication.)
- Dissemination of Work. (Presentation of results in writing, oral presentations, and informal project discussion with others in and outside of the lab is clear, correct, and concise.)
- Understanding of the Science. (Scholar understands the science and statistics underlying the research. Scholar is able to identify and seek out answers to gaps in their knowledge.)
- Ability to Work Independently and with Others. (Scholar has good time management skills and is able to set realistic timelines that they can complete. Scholar is professional when collaborating with others.)

Example self-evaluation questions:
- What do you consider your strengths?
- What do you consider your weaknesses?
- Which, if any, weaknesses do you want to improve?
- What were specific accomplishments for you for this rotation or research experience?
- What are your goals for your next rotation or research experience?
- What do you hope to improve in your next rotation or research experience?
- How do you think the research team or research experience could improve?*
- What can I (your mentor) do to help you reach your goals?*
(*) may receive more honest feedback through an anonymous survey if possible

***Set reasonable expectations (and remember, most undergraduate research is training).***
Mentees and mentors can be surprised at how much of the trainee's time is taken up by training rather than research. Keeping in mind that training and learning make up the vast majority of undergraduate research experiences, especially at the beginning, can help increase enjoyment and ensure the setting of realistic goals **(Figure 2)**. Consider five-levels of research experiences: level 1 (entry), level 2 (beginner), level 3 (moderate), level 4 (advanced), level 5 (expert). Most undergraduate researchers start at level one, entry, with no to very little research experience. Trainees in this category spend ~90% of their time on learning and training. After some initial training, students progress to level two (~75% training/25% research) and then to level three, moderate (~50% training/50% research). Most undergraduate researchers will only get to level 2 or level 3 before graduating. Setting the expectation early on that most of the undergraduate researcher's time will be training can help the scholar feel successful in the experience.

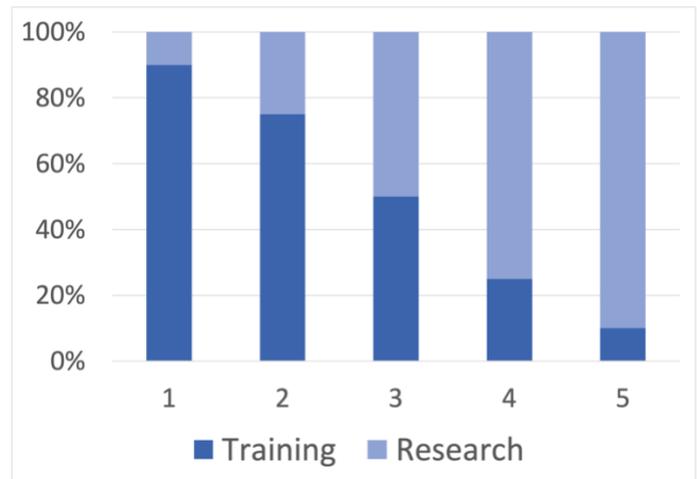

**Figure 2. Proportion of time for research and training.** The amount of time research assistants spend on training and research will vary by level. *Level 1* trainees have limited to no prior experience research and will spend ~90% and 10% of time training completing research respectively. These trainees are often high-school, undergraduate, and MS students. *Level 2* trainees have little research experience and will spend ~75% of time training and ~25% of time completing research. These trainees may be high-school, undergraduate, and first year graduate students. *Level 3* trainees have some prior experience research and will spend ~50% of time training and ~50% completing research. These trainees can be experienced undergraduate researchers or graduate students. *Level 4* trainees have a lot of research experience and will spend ~25% of time training and ~75% of time completing research. These trainees are often advanced graduate students. *Level 5* trainees have considerable research experience and will spend ~10% of time training and ~90% of time completing research. These trainees are often sometimes advanced graduate students although they are usually post-doctoral fellows.

***Provide a structured and supportive start-up ramp.***
Most undergraduate researchers will need significant training. Additionally, removing barriers to entry will necessitate extra training either prior or in-parallel with completing research. A start-up process can provide structure and set expectations for both trainees and faculty (**Figure 3**). The start-up process can include a start-up to do list that can be mostly completed independently with online trainings and readings. Depending on a student's background this might include coding basics in R or Python, human subjects training, reading about the application area (e.g., human genetics) and previous laboratory papers. To support meeting the team, the start-up to do list can also include individual meetings with team members and a short paragraph write-up of their research. Once the start-up to do list is complete, students can progress through replication of an existing project enabling them to learn and apply necessary skills in a situation where the result is known. Then scholars proceed to shadowing (i.e., working with another person on a project), taking on a small independent project, and finally completing a larger project. Each phase may take anywhere from 3 weeks to 3 months or longer depending on the research time available, the background of the student, and the research scope.

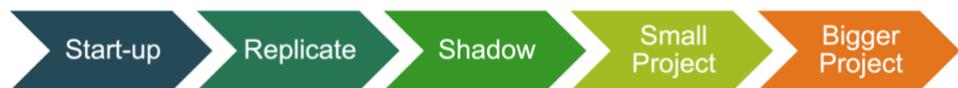

**Figure 3. Structured start-up ramp.** Starting with a ***start-up*** to do list consisting of online trainings and readings that can be accomplished mostly independently, scholars progress to ***replicate*** an existing small result within an ongoing project. Replication helps trainees learn aspects of the research process (e.g., coding, data wrangling, installing and using software) in a supportive structure where the final result is known. Scholars then become more immersed in a project by ***shadowing*** an ongoing project from a team member coding in parallel and providing code review. Finally, scholars progress to completing a ***small project*** that is a subset of an ongoing project and onto a ***bigger projects*** where they can take on a more independent role. Each phase of the start-up ramp can take between 3 weeks to 3 months or longer depending on the background of the trainee, the time commitment available, and the needs of the project.

*Include trainees in real, significant, but not time-sensitive research.*
Every now and then, a student will start in a research group with their own well formulated, attainable, and significant research question. However, it is far more common that students arrive with an interest but little idea of what questions are relevant and, importantly, are attainable within a short time-frame. Starting with a faculty driven research question helps introduce the student to the research process with an attainable and important research goal. This can prove to be more motivating and exciting for a trainee, especially as they learn about the significance of the work. Additionally, this can align well the faculty's research making better use of the faculty's time. Importantly, avoid essential or time sensitive projects for entry or even moderate level (level 1 and 2) trainees as the research vs training expectations are out of line with time sensitive work and can result in a sense of failure for both trainee and mentor if the deadline is not met.

*Be realistic and positive.*
Remind yourself and your trainee that research takes time and that obstacles and setbacks are common. Normalize failure and the success of identifying mistakes. This is especially important for trainees new to research. Set and remind students of the expectation to take care of themselves. Ultimately, our goal is to introduce students to the amazing nature of research, and data science. Regardless of whether the trainees go on in a research centered career, a positive training experience can help them succeed in a variety of industry and other careers and may lead to other connections as they speak about their positive experience.

*Provide many opportunities to disseminate research.*
Dissemination of research increases understanding and motivation of science. Dissemination opportunities can include research team meetings, the department, university, and local or national/international conferences. Many organizations provide reduced or free student memberships (e.g., American Statistical Association [ASA], International Biometrics Society [IBS], Society for Industrial and Applied Mathematics [SIAM]) and small, local meetings that are highly supportive of student presentations. Starting with research group meetings can provide experience in a supportive environment. Practice different formats of dissemination including posters, presentations, elevator pitches, abstract writing, and manuscript writing.

*Teach soft skills and the hidden curriculum.*
As important as learning about the technical aspects of research is learning the "soft-skills" needed for a successful career in industry and academia. This training can occur within your research team meetings, at the college or university level, or through professional organizations (e.g., ASA and CUR). Teaching soft skills can be especially important for first-generation students and trainees from different cultures. Topics can include how to apply for jobs, graduate school, and grants (NSF & NIH fellowships); managing your mentor/boss; how to write emails so people respond; how to be a strong self-advocate; how to network. A piece of the hidden curriculum that is important to emphasize especially for students new to knowledge driven work, is that work (all of it) deserves to be paid. Sometimes trainees who are being paid by the hour will not count time they spend learning or in tasks that they deem are not directly related to research (e.g., writing emails). Being clear that any time spent related to the project is part of their paid time helps set standard for your research team, and importantly for their future work.

## Retaining
*Remember, research training is a marathon, not a sprint.*
Trainees may be excited, and perhaps a little nervous, to start a research experience. In addition to setting clear expectations as outlined above, reminding trainees that both health and well-being and and school come first is helpful. While research is exciting and important, without first maintaining health/well-being and success in school, successful undergraduate research is simply not possible.

*Create a supportive mentoring structure.*
While a faculty mentor is an important source of research mentorship providing guidance on how to complete impactful and attainable research, peer and near-peer mentors provide additional sources of support[10]. Peer and near-peer mentors can provide advice on the next steps in a research career and words of wisdom or support from someone who has recently been through a similar experience. Additional mentors can also provide more regular check-ins with mentees that may not be possible for one faculty member. In addition to having mentors, undergraduate researchers can be near-peer mentors for others, such as high-school or community college students. Being a mentor themselves can help trainees identify successful and unsuccessful parts of their path while being an example of a possible path for others. Near-peer mentoring has been shown to be highly beneficial for both mentees and mentors and especially for those from underrepresented groups[11-13]. This broad mentor and mentee structure can create a mentoring network (**Figure 4**) that can help trainees build a community and sense of belonging.

*Create an inclusive and supportive community.*
Getting to know your team and helping the team get to know each other can help foster a sense of community that can result in improved cohesion, collaboration, and retention. Fostering a sense of community can start small such as making time for check ins at the beginning of each team meeting and 1to1. These can be in the form of "what happened and what's coming up" or "updates/struggles/shoutouts" where team members provide a short update, a struggle, and a shoutout to a member of the team – the updates/struggles/shoutouts can be personal or about work. Faculty sharing, both personal and professional, can help normalize struggles. Semester group activities such as hikes or pumpkin carving can help foster a sense of community as well. Importantly, all updates and activities are optional.

*Normalize the research process.*
Failure, set-backs, imposter syndrome are all quite real in research. Openly talking about these in team meetings and 1to1s can help normalize the research process and help trainees not feel like failures when a setback has occurred.

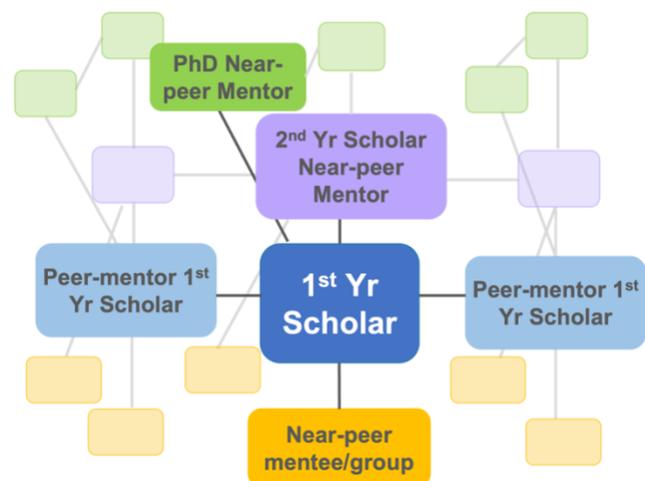

**Figure 4. Mentoring Network.** By employing near-peer and peer mentoring, scholars create a network and community; building their identity and belonging as researchers.

*Foster identities as data scientists and researchers*
A sense of belonging has been shown to increase completion of degrees and retention in programs[3,14-16]. Thus, a foundational component of undergraduate research is to help scholars develop an identity and a belonging as a researcher and data scientist. Imposter syndrome is real and, often, stronger for trainees from underrepresented backgrounds and can be especially magnified when trainees do not see people from similar backgrounds and with similar experiences. In addition to helping trainees form a mentoring network and creating a supportive team community, connecting students to groups where they can see and get to know others who have taken similar paths can be especially valuable. Societies such as the Association for Women in Mathematics (https://awm-math.org/), Society for Advancing Chicanos/Hispanics & Native Americans in Science (https://www.sacnas.org/), and CUR (https://www.cur.org/) can be helpful in supporting undergraduate researchers in forming their identities as scientists.

## Final Thoughts

*Lack of funding for research experiences is an equity issue.*
Lack of funding for research experiences creates an inequitable barrier to research experiences and subsequently graduate school and research careers. Solutions to funding include individual level programs such as work-study or University/College-level student grants and larger programs through NIH and NSF. Additionally, near-peer mentors deserve to be paid for their mentoring time.

*Keep iterating.*
The best structure and support for a person and a team will change. Continuing to iterate will help the team evolve to meet the needs of its members. Use feedback from the trainees to help iterate.

*Make time for mentor training.*
Training is essential to helping support mentor development. Training can take the form small groups of practice with other faculty members to mentorship seminars to in person or online workshops such as Entering Mentoring[17]. CUR has numerous training resources on its website.

## In Closing

Mentoring undergraduates in research can be extremely rewarding and, when done thoughtfully, can align with current research, teaching, and service roles. When successful, mentors get to train the next generation of researchers while also advancing their research area and trainees get to learn about possible new areas of interest and careers as data scientists.

# APPENDIX

## **Hendricks Lab Expectations**

*Welcome* to the Hendricks lab. Here, we complete world-class applied and methodological statistics research using genetics and 'omics to dissect the complex nature of human diseases and traits. We work in a collaborative setting with people from a variety of backgrounds and education levels. We are always learning, improving, and pushing ourselves and others to be our best. In doing so, we produce first-class research for the broader community and train the next generation of genomics and health data scientists.

*My Responsibilities.*
- I will do my best to help you become an exceptional researcher and to help you achieve your goals whether they be in academia or industry or other.
- I will provide timely, honest, and constructive feedback and thoughtful advice.
- I will do my best to provide a supportive, productive, and collaborative work environment.
- I will strive to set high and achievable research goals pushing the team to always improve.

*Team Expectations.*
- Take care of your mental, emotional, and physical health – if you are burnt out, you cannot do good work
- Make time for life to happen
- Commit to first-class rigorous and reproducible research
- Respect yourself and others
- Communicate proactively
- Be a contributing and collaborative member of the team
    - Work through issues or conflicts with other team members should they arise
    - Respond to emails or slack messages in a timely manner (e.g. ~24 hours) unless on vacation; expected responses include confirmation of receiving the email
- Help make our group inclusive and welcoming to all people (e.g., all races, ethnicities, genders, sexual orientation, age, background, religion)

*Lab Meetings.* Lab meetings will be weekly or biweekly. It is the expectation that research assistants attend each lab meeting unless otherwise determined. During lab meetings, we will connect as a team, set goals and timelines, and practice communicating our research. Please contact Audrey or your graduate student co-mentor if you have a conflict and will be unable to make a meeting or 1to1; at least 24 hours' notice is expected when possible.

*Expected Time Commitment.* A sufficient and regular time commitment to the research in the lab is necessary both to move the science forward and for the research assistant to have a beneficial research experience. As such, unless otherwise discussed with Dr. Hendricks, research assistants are expected to schedule at least 12 hours/week to work on the project of which at least 8 should be "in person" time available in person or for zoom and teams as needed. The number of hours committed to research will increase for Masters and PhD students (>20 hrs/week). Health and wellbeing are essential. As such, it is expected that research assistants will take off ~10% of their time each month.

*Office:* There is an office available for use by all researchers in our team. Please keep the office clean for the next people to use. The office should only be used by members of our research team. The office is on the math/stats floor room XXXX.

*Regular check-ins.* Research assistants will attend weekly or bi-weekly check ins with Audrey and the graduate student co-mentor.  Assistants will learn to lead their one-to-ones (i.e., managing their manager) guiding the flow of the meeting.
- an update (preferably in writing) on what they did last week
- what the plans are for the upcoming week
- questions or discussion topics to advance their research. It is expected that assistants will have at least one question/discussion topic.

*Work time.* It is expected that research assistants will coordinate with others working on similar projects. Assistants will check teams and email to see what others are working on and will communicate often with others. **For the fall semester, the default expectation is to be in person on Fridays.**

*Health and Wellbeing.* Health and wellbeing are absolutely essential to a productive team and, most importantly, researcher. Health and wellbeing includes physical, mental, and emotional well-being. Remember that sufficient sleep is necessary too! It is my expectation that research assistants will seek help and resources as needed including talking with me, and others in the team as comfortable. Remember to keep time in your life for life to happen.  And, remember you can always readjust!  CU Denver has free counseling and other resources (http://www.ucdenver.edu/life/services/counseling-center/about/Pages/default.aspx?gclid=CjwKCAjwqZPrBRBnEiwAmNJsNv60cgOz09d3OdftzaH734eIdMLvSxpVmYAfcBqmNKEDp8DX-uV2-RoCrgIQAvD_BwE). I and the rest of the team are committed to supporting the health and wellbeing of our team members. We will identify the best way forward to accommodate people.

*Seminars.* As part of this research team, it is encouraged that you attend some (not all) weekly seminars including
- Mathematical and Statistical Sciences Seminars: Mondays from 11-12:15. See here for calendar information: https://clas.ucdenver.edu/mathematical-and-statistical-sciences/event-calendar.
- Human Medical Genetics and Genomics Program Seminars: https://www.cuanschutz.edu/graduate-programs/human-medical-genetics-and-genomics/events

I have read, understand, and agree to the Hendricks Lab Expectations written above.

______________________________________                        _______________
        Name                                                                                                           date

# Hendricks Team meeting guidelines

Below are pre-, during, post- meeting expectations to help us have productive and efficient meetings. Each meeting will have an organizer and a note taker. When applying instructions in this document, replace text in ⟨angle brackets⟩ with whatever is described by the bracketed phrase. Please email me with any questions or concerns. For suggested modifications or clarifications, please add comments to this document.

Table 1. Meeting types, roles, and responsibilities

|  | Title (for documents) | Organizer | Note Taker | Send Agenda/ Paper | Upload Notes/ Presentation |
|---|---|---|---|---|---|
| **1:1** | ⟨your name⟩[a] 1:1 | Team member | Team member | 24 h before | 24 h after |
| **2:1** | ⟨your name⟩ 2:1 | Junior team member | Junior team member | 24 h before | 24 h after |
| **Lab meeting: Journal club** | Lab meeting: Journal: ⟨paper title⟩⟨date⟩ | presenter | – | 1 w before | 24 h after |
| **Lab meeting: Soft Skills** | Lab meeting: Tech: ⟨tech title⟩⟨date⟩ | presenter | – | – | 24 h after |
| **Lab meeting: Research** | Lab meeting: ⟨your name⟩⟨date⟩: Research | presenter | presenter | – | 24 h after |
| **Supervisory/ dissertation committee** | Supervisory committee: ⟨your name⟩ ⟨date⟩ | presenter | presenter | 1 w before | 24 h after |
| **Other meeting** | ⟨meeting name⟩ ⟨date⟩ | Senior team member [b] | Junior team member | 1 w before | 24 h after |

[a] Use date format: yyyy-mm-dd
[b] not Audrey

# Meeting roles

Each meeting has an organizer, who posts and emails the agenda. Most meetings have a note taker.

**Organizer responsibilities**
1. Communicate the agenda **in advance**. Rationale: ensure everyone is fully prepared for the meeting. Meetings should be used for things that are more complex and nuanced than those that can be dealt with over email.

2. Agenda content
    a. Agenda: a numbered list of items or questions to discuss, each one no more than one line.
    b. Short explanatory notes can be later on in the document, if required.

3. **Posting** the agenda:
    a. Creating agenda for a new meeting or series:
        i. Create a OneDrive Word Doc with title: "Agenda: ⟨title from Table 1⟩" under your OneDrive folder (for 1:1 or 1:2) or the topic area for research meetings
        ii. For a meeting series (e.g., 1to1s), have one agenda and add the new date and agenda to the top of the document
        iii. Add the link to your agenda to your lab notebook (e.g., OneDrive folder, goals board, word document, PowerPoint).
    b. Converting a one-off meeting to a series:

i.   Re-use the Doc created for the first meeting, changing its title to be the series.
            ii.  Put new material at the top of the document
      c. Type the agenda content into the Google Doc (see example at end of document).

   4. **Sending** the agenda:
      a. Create an email with subject: "Agenda: ⟨title from Table 1⟩, ⟨date⟩"
            i.   ⟨date⟩ is ISO 8601 format YYYY-MM-DD. For example: 2020-03-23
      b. Copy the contents of the OneDrive Doc into the text of an email.
      c. At the bottom, paste the URL of the agenda on OneDrive Doc (even if you have sent out this URL before).
      d. Send to all participants.
      e. Please email all participants any later updates to the agenda.

**Note taker responsibilities**
   1. Edit the Agenda Google Doc to include minutes directly underneath individual agenda items.
      a. Minutes content
            i.   Record any decisions made in the meeting.
            ii.  Action items. For any actions discussed that need execution by meeting participants, put "**ACTION**: ⟨responsible person's name⟩: ⟨action to be taken⟩" with **"ACTION:"** in bold and all caps. This includes any actions that need to be taken by Michael or other faculty.
            iii. Usually it is unnecessary to take detailed notes, outside the categories of (i) records of decision and (ii) action items.
   2. **Sending** the minutes:
      a. As soon as possible after the meeting (within 24 hours), copy the contents of the agenda with notes into the text of an email and send it to all participants with the URL.
      b. Copy the notes into your lab notebook for a durable record.

## Summaries for 1:1 and 2:1 meetings
Only for the last scheduled 1:1 and 2:1 meeting of the week, beneath the agenda, briefly summarize your progress and priorities. This should take no more than 15 min to write, and no more than 5 min to read. On weeks when you do not have a meeting with Audrey, post and send her your summary instead of an agenda.

Include these sections:
   1. ***To be discussed.*** Agenda for the meeting.
   2. ***Accomplishments for the week***. List completed activities and notable accomplishments. In general, what is working? What is your current situation?
   3. ***Priorities for next week.***
   4. ***Challenges and roadblocks.*** Describe potential challenges that have impeded or may impede your intended tasks and goals.
   5. ***Lessons learned and opportunities for improvement.*** List any area that might benefit from improvement within our control; problems you are trying to solve; lessons recently learned or relearned.
   6. ***Literature***. Include an interesting paper since the last meeting and a brief summary of why you chose this paper
   7. ***Other activities.*** List other activities that are related to your work but are outside of research that you accomplished this week. This will help you appreciate and track all that you do.
   8. ***Date and location of Next Meeting***. Include the date and location of our next meeting so we can make sure we're on the same calendar.

# Example (you can copy this as a template)

## 1:1 Michelle and Audrey, 2020-05-14

### To be discussed
1. Hexbin plots
2. Update to correlation plots

### Accomplishments
1. Parallelized the adjusting methylation code—currently running
2. Redid the correlation plots removing windows with mappability score <0.5
3. Did the hexbin plots looking at relationship between fmol and standard deviation
4. Batch 2 is now UMI deduplicated

### Priorities for next week
1. Finish batch analysis, add to presentation
2. Make changes to presentation based of lab meeting feedback and practice presentation
3. Finish touching up manuscript draft

### Challenges and roadblocks
1. Most of batch analysis is contingent on the job scheduler and is taking time
2. I've had some issues with methylation adjustment and resulting files being corrupted which is pushing me back. Correcting this now

### Lessons learned and opportunities for improvement
1. Learned I need more memory to run files of the size I have. Going to learn how to migrate things over h4h and run on the new cluster.

### Literature (an interesting paper since the last meeting)
1. Wojcik GL, Murphy J, Edelson JL, Gignoux CR, Ioannidis AG, Manning A, Rivas MA, Buyske S, Hendricks AE. Opportunities and challenges for the use of common controls in sequencing studies. Nat Rev Genet. 2022 Nov;23(11):665-679. doi: 10.1038/s41576-022-00487-4. Epub 2022 May 17. PMID: 35581355; PMCID: PMC9765323.
   - https://www.ncbi.nlm.nih.gov/pmc/articles/PMC9765323/
   - Summarizes methods, datasets, and other considerations when using common controls. May be useful to cite for the ProxECAT v2 manuscript.

### Other activities (other activities such as service, teaching)
- Attended SSGG meeting as committee member
- Reviewed a manuscript for Nature Genetics

### Date and location of Next Meeting: Tuesday, January, 3 2022 at 11:30 in person at AMC

# Goal Setting

What are your academic goals for this semester?

What are your research goals for this semester?

Do you have any additional goals for this semester?  If so, what are they?

What are your 1-, 3-, and 5-year rough goals? (feel free to copy or change slightly from last time)

What can I do to help you achieve your goals?

What are your concerns for next semester?

What can I do to help alleviate your concerns?

# Student Self-Reflection.

How do you do with time and expectation management and productivity?

What do you consider your strengths?

What do you consider your weaknesses?

Which of your weaknesses do you want to focus on improving first? What tangible things will you try to work on this/these weakness?

What barriers to success are standing in your way?

**This Semester _______**

Specific accomplishments for this semester

What went well for you this semester

What went well for the team this semester

# Hendricks Research Group Progress Report (undergraduate)

NAME. Walter Baer
DATE. 12/15/22

Summary: Below is a summary of the progress of four general areas for the student. The scale ranges from 1 to 5 with 1 being the lowest value and 5 being the highest. 1-serious concerns, 2-poor, 3-needs improvement, 4-good, 5-excellent. Students are expected to strive for 4s and 5s across all areas. To continue with the research group, students are expected to maintain scores above 3. Scores at a 3 or below for two or more semesters may not be invited to continue with the research group.

**Fall 2022.** Summary of feedback.

| | |
|---|---|
| **General**. Ontime updates and progress reports. Clear reporting of current work and goals. Accurate tracking of time and progress. Responsive in a timely manner to emails and other communication.<br><br>Comments | 4.5 |
| **Dissemination of Work**. Presentation of results in writing, oral presentations, and informal project discussion with others in and outside of the lab is clear, correct, and concise.<br><br>Comments | 4 |
| **Understanding of the Science**. Student understands the science and mathematics underlying the research. Student is able to identify and seek out answers to gaps in knowledge.<br><br>Comments | 3.5 |
| **Ability to Work Independently and with Others**. Student has good time management setting realistic timelines that they can complete. Student works well with others including finding time to meet others as needed.<br><br>Comments | 4 |
| **Technology**. Student is able to learn and use the necessary technology and programming languages. Student seeks out additional resources and help as needed. Student arrives at a reasonable balance between trying to learn on their own and asking questions/seeking help.<br><br>Comments | 5 |

**Near-peer Mentor feedback:** xxx